\documentclass[12pt]{article}
\pdfoutput=1
\usepackage{amsmath,amssymb,amsfonts,amsthm,bm,bbm,cancel,wasysym}
\usepackage{epsfig,graphics,graphicx,epstopdf,caption,subcaption}
\usepackage[numbers,sort&compress]{natbib}
\graphicspath{{Charts/}}
\usepackage{array,booktabs,colortbl,colordvi,multirow}
\usepackage{colordvi,color,xcolor}
\usepackage{hyperref}
\usepackage{rotating}
\usepackage{comment}
\usepackage{feynmf}

\begin{document}

\begin{center}
{\Large {\bf Freeze-in Dark Matter in EDGES 21-cm Signal}}\\

\vspace*{0.75cm}

{Shengyu Wu$^{1}$, Shuai Xu$^{2}$ and Sibo Zheng$^{1}$}

\vspace{0.5cm}
{$^{1}$Department of Physics, Chongqing University, Chongqing 401331, China\\
$^{2}$School of Physics and Telecommunications Engineering, Zhoukou Normal University, Henan 466001, China}
\end{center}
\vspace{.5cm}

\begin{abstract}
 
\noindent 
The first measurement on temperature of hydrogen 21-cm signal reported by EDGES
strongly favors Coulomb-like interaction between freeze-in dark matter and baryon fluid.
We investigate such dark matter both in one- and two-component context,
with the light force carrier(s) essential for the Coulomb-like interaction not being photon.
Using a conversion of cross sections used by relevant experiments and Boltzmann equations to encode effects of the dark matter-baryon interaction, 
we show that both cases are robustly excluded by the stringent stellar cooling bounds in the sub-GeV dark matter mass range.
The exclusion of one-component case applies to simplified freeze-in dark matter with the light force carrier as dark photon, gauged $B-L$, $L_{e}-L_{\mu}$,$L_{e}-L_{\tau}$ or axion-like particle,
while the exclusion of two-component case applies to simplified freeze-in dark matter with the two light force carriers as two axion-like particles coupled to standard model quarks and leptons respectively.
\end{abstract}

\renewcommand{\thefootnote}{\arabic{footnote}}
\setcounter{footnote}{0}
\thispagestyle{empty}
\vfill
\newpage
\setcounter{page}{1}

\tableofcontents
\section{Introduction}
\label{introduction}
Cosmological surveys in decades enable us to draw a picture of modern cosmology based upon $\Lambda$CDM baseline model, except with several fundamental puzzles such as the nature of dark matter (DM).
Among various efforts to detect DM,
21-cm cosmology \cite{Pritchard:2011xb} as a probe of spin flipping of ground-state hydrogen atom of baryon gas during dark ages,
provides a new window to search for dark matter (DM) in a rather low velocity region.
Since brightness temperature $T_{21}$ of hydrogen 21-cm line 
is tied to baryon temperature $T_b$,
a measurement on $T_{21}$ sheds light on DM-baryon interaction \cite{Munoz:2015bca,Tashiro:2014tsa,Dvorkin:2013cea}  which  can affect $T_b$.
Recently, the EDGES experiment reported the first measurement on the sky-averaged value \cite{Bowman:2018yin}
\begin{eqnarray}{\label{Edata}}
\left<T_{21}\right>= -500^{+ 200}_{- 500}~\rm{mK},
\end{eqnarray}
at redshift $z\approx 17$,
which deviates from the prediction of $\Lambda$CDM with a significance of $\sim 3.8~\sigma$.\footnote{The signal significance is disputed by SARAS3 experiment \cite{Singh:2022ivh} and a reanalysis of the EDGES signal \cite{Hills:2018vyr}.}
As this signal strongly favors DM-baryon interaction, 
it is natural to ask what kinds of DM model, with what kind of DM-baryon interaction, 
within which DM mass range, can cool down the baryon gas to explain the EDGES data.

Several works have advocated Coulomb-like interaction between the DM and baryon as a solution \cite{Munoz:2018pzp, Barkana:2018lgd, Berlin:2018sjs,Barkana:2018qrx,Slatyer:2018aqg,Kovetz:2018zan,Boddy:2018wzy,Creque-Sarbinowski:2019mcm, Liu:2019knx,Aboubrahim:2021ohe,Li:2021kso} to the observed signal in the literature.
If so, a massless or a light force carrier $\omega$ is essential, 
implying that such DM is actually freeze-in.\footnote{Because the couplings of a light force carrier to electron and proton have to be far less than unity, otherwise they have been excluded by lepton and/or hadron colliders respectively. If so, the feeble interactions are unable to keep the DM in thermal equilibrium with the SM thermal bath in early universe. Rather, they allow it to freeze-in \cite{Hall:2009bx}, leading to freeze-in DM.}
The freeze-in DM-baryon elastic scattering cross section scales as $\sim \upsilon^{-4}_{\rm{rel}}$, 
with $\upsilon_{\rm{rel}}$ the relative velocity of interacting particles.
This velocity-dependent behavior significantly amplifies the effect on $T_b$ at the low velocities such as $\upsilon_{\rm{rel}}\sim 10^{-6}$ at the redshift $z\approx 17$, compared to those at relatively higher $\upsilon_{\rm{rel}}$ such as DM direct detection experiments (with $\upsilon_{\rm{rel}}\sim 10^{-3}$) or early Universe (with $\upsilon_{\rm{rel}}\sim 0.1-1$).\footnote{Throughout the paper, 
 velocity is in units of light speed $c$ if not mentioned.}
In other words, Coulomb-like interaction naturally provides a large gap between the cross sections related to the signal and the aforementioned constraints among others.
Nevertheless, those previous studies indicate that such a gap seems still inadequate in simple DM models.

\begin{figure}
\centering
\includegraphics[width=13cm,height=9cm]{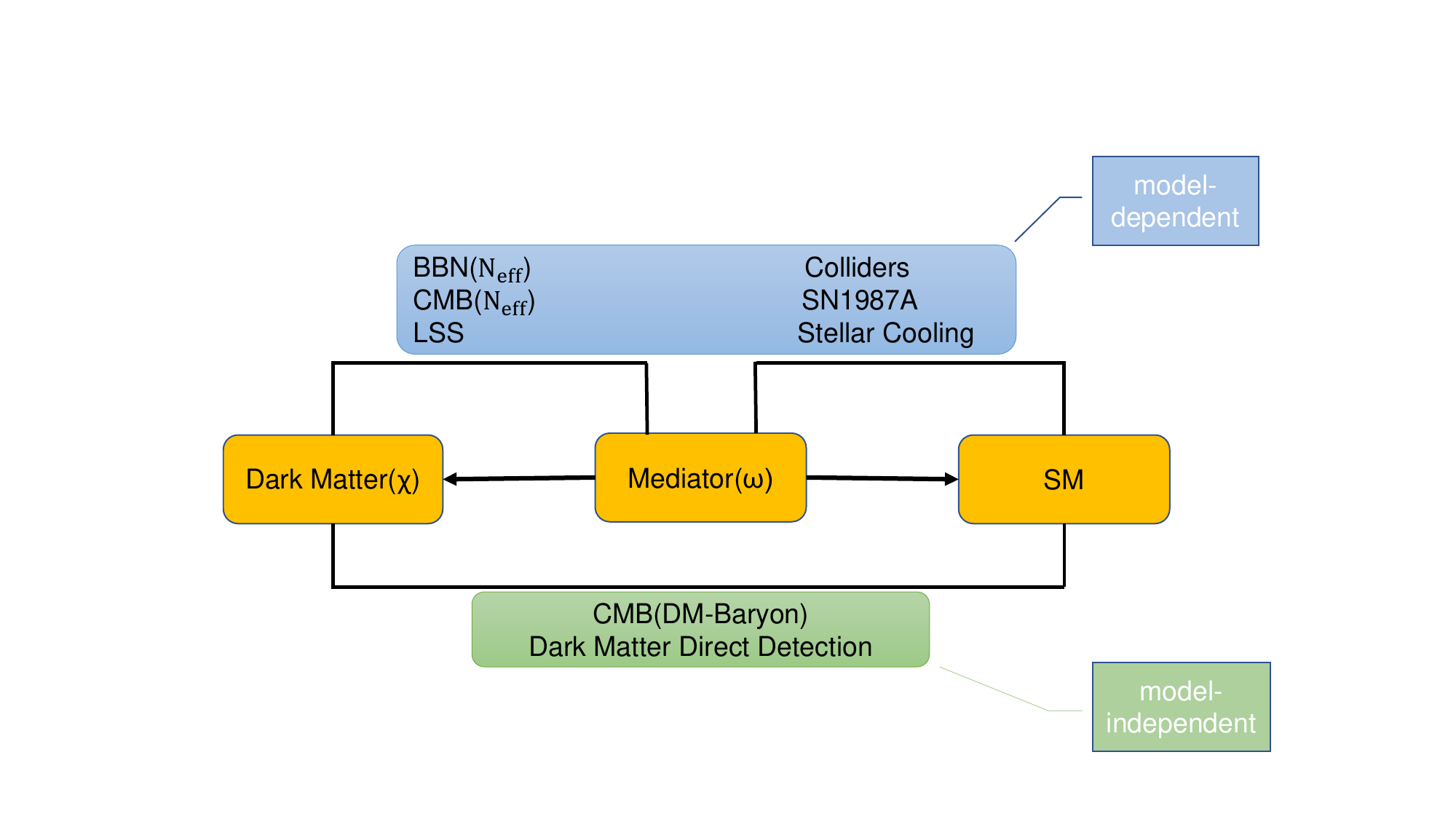}
\centering
\caption{Overview of the most relevant constraints on the freeze-in DM with eV-keV scale force carrier(s) into model-dependent (top) and model-independent (bottom) ones. 
As we will explain, the most stringent stellar cooling bounds can be promoted to be mode-independent.}
\label{sketch}
\end{figure}

In this study we instead investigate such freeze-in  DM scenario through a model-independent survey. 
To be concrete, we focus on the freeze-in DM interacting with the charged particles of the baryon gas, 
namely electron and proton in the standard model (SM).
For the light force carrier mass heavier than eV scale as we will consider, 
the most relevant constraints from DM direct detections such as DM-electron (e) scattering \cite{Essig:2012yx, Essig:2017kqs, DarkSide:2018ppu, SENSEI:2019ibb, DAMIC:2019dcn, SENSEI:2020dpa,PandaX-II:2021nsg, XENON:2021myl} 
and DM-proton (p) scattering \cite{LUX:2017ree,PandaX-II:2018xpz, XENON:2018voc, XENON:2019zpr, Akimov:2021yeu, Adhikari:2021kwv},
early cosmology such as big bang nucleosynthesis (BBN) \cite{Vogel:2013raa} and cosmic microwave background (CMB) \cite{Boehm:2013jpa, Slatyer:2018aqg, Dvorkin:2013cea, dePutter:2018xte, Xu:2018efh},
stellar cooling \cite{Davidson:2000hf, Chang:2018rso,Hardy:2016kme, An:2014twa}, large-scale structure (LSS) \cite{Tulin:2013teo,Tulin:2017ara} as well as colliders \cite{Alekhin:2015byh, Beacham:2019nyx} can be divided into model-dependent and model-independent catalog, see Fig.\ref{sketch} for an overview.\footnote{Below eV scale, the stellar cooling bounds are taken over by the stronger 5th force experiments \cite{Adelberger:2003zx, Salumbides:2013dua}.}
Among these constraints, the stellar cooling bounds are the most stringent in the parameter regions where they are present.
Thanks to that the stellar cooling bounds can be promoted to model-independent ones,
we are able to find out whether viable parameter space exists after the most important model-independent constraints are imposed. 
A surviving region, if any, is useful for DM model builders not familiar with the 21-cm cosmology.

This paper has the following structure.
In Sec.\ref{crosssection}, 
we introduce two different descriptions of the scattering cross sections used by the EDGES 21-cm signal analysis and the DM direct detection experiments.
Sec.\ref{Bequations} is devoted to Boltzmann equations which govern the evolution of the temperatures of both DM and baryon fluid after kinetic decoupling,
where the one- and two-component DM are considered in  Sec.\ref{B1} and Sec.\ref{B2} respectively.
By numerically solving the Boltzmann equations, 
we analyze the parameter space with respect to the EDGES signal in the sub-GeV DM mass range for the one- and two-component case in Sec.\ref{space1} and Sec.\ref{space2} respectively, and point out physical implications to simplified freeze-in DM models.
Finally, we conclude in Sec.\ref{con}.

\section{Scattering Cross Sections}
\label{crosssection}
The cross sections $\hat{\sigma}^{I}$ used by the EDGES signal analysis are given by
\begin{equation}{\label{comp}}
\sigma^{I}_{T}=\hat{\sigma}^{I}\upsilon^{-4}_{\rm{rel}},
\end{equation}
where $I=\{e, p\}$ refers to a component of baryon gas contributing to the DM-baryon scattering,
and $\sigma^{I}_{T}$ is the so-called momentum-transfer cross section defined as \cite{Tulin:2013teo}
\begin{eqnarray}{\label{cs}}
\sigma^{I}_{T}=\int d\Omega (1-\cos\theta)\frac{d\sigma^{I}}{d\Omega},
\end{eqnarray}
with $d\sigma^{I}/d\Omega$  the differential cross section and $\theta$ the scattering angle.

As noted in earlier studies e.g, \cite{Barkana:2018qrx},
$\hat{\sigma}^{I}$ are different from the cross sections $\bar{\sigma}^{I}$ used by the DM direct detection experiments defined as
\begin{eqnarray}{\label{cse}}
\frac{d\sigma^{I}}{d\Omega}=\frac{\bar{\sigma}^{I}}{4\pi}\bigg| F_{\chi}(q^{2})\bigg|^{2} \bigg| f(q^{2})\bigg|^{2},
\end{eqnarray}
where $F_{\chi}(q^{2})\sim 1/q^{2}$ is the DM form factor and $f(q^{2})\approx 1$ is the target form factor respectively.
Substituting Eq.(\ref{cse}) into Eq.(\ref{cs}) gives \cite{Barkana:2018qrx}
\begin{eqnarray}{\label{transf}}
\bar{\sigma}^{I}\approx 8\xi^{-1}_{I}\left(\frac{\mu_{I}}{q_I}\right)^{4}\hat{\sigma}^{I},
\end{eqnarray}
where $\mu_{I}$ is the DM-$I$ reduced mass, $q_{I}$ is the typical moment transfer in relevant scattering process, and $\xi_{I}$ is a logarithm term
\begin{eqnarray}{\label{log}}
\xi_{I}\approx\log\left(\frac{4\mu^{2}_{I}\upsilon^{2}_{\rm{rel}}}{m_{\omega}^{2}}\right).
\end{eqnarray}
with $m_{\omega}$ the force carrier mass. 

In order to compare with DM direct detection limits, 
we have to convert $\hat{\sigma}^{I}$ obtained from the EDGES signal to $\bar{\sigma}^{I}$ in terms of Eq.(\ref{transf}),
where the explicit values of $q_{I}$ and $\xi_{I}$ are DM experiment-relevant. 
In the DM mass range $m_{\chi}\sim 1-10^{3}$ MeV,
the most stringent limits can be placed as follows.
For the DM-e cross section $\bar{\sigma}^e$,
we pay attention to both SENSEI \cite{SENSEI:2020dpa} and XENON \cite{XENON:2021myl} limits, 
which are able to constrain most of the mass ranges. 
In these experiments $q_{e}\approx \alpha m_{e}$ as in \cite{Barkana:2018qrx} with $\alpha$ the fine structure constant, 
and the value of $\xi_{I}$ is determined by $m_{\omega}$ required by $m_{\omega}\leq 10^{-6}\mu_{I}$ in the red shift region $z \sim 10-10^{3}$ and the relative velocity $\upsilon_{\rm{rel}}\sim 10^{-3}$ at these DM direct detection experiments.
For the DM-p cross section $\bar{\sigma}^p$ which is spin-dependent as we will assume, 
we consider XENON1T \cite{XENON:2019zpr}  limit to constrain $m_{\chi}$ down to $\sim$ 80 MeV.
At the XENON1T experiment, the moment transfer $q_{p}\approx \sqrt{2m_{p}E_{R}}\sim 1.0-2.0$ MeV \cite{XENON:2019zpr} with $E_{R}\sim 1-2$ keV the recoil energy.

\section{Boltzmann Equations}
\label{Bequations}
In this section we derive the Boltzmann equations which govern the evolution of temperatures of both DM and baryon gas as perfect fluids after the kinetic decoupling. 
The DM-baryon interaction, which is non-relativistic, yields two main effects \cite{Munoz:2015bca,Tashiro:2014tsa,Dvorkin:2013cea}: a transfer of heating and a change of relative velocity between the two fluids.

\subsection{One-component DM}
\label{B1}
With DM being a single component, 
the transfer of heating is described by the heating terms $Q_{b, \chi}$ which respect energy conservation, 
while the change of relative velocity between the DM and baryon fluid is characterized by the drag term $D$. 
In the situation where the DM particle $\chi$ simultaneously interacts with the different components $I$ of the baryon fluid,
they are given by
\begin{eqnarray}{\label{DQ1}}
D&=&-\dot{V}_{\chi b}=\sum_{I}\frac{(\rho_{\chi}+\rho_{b})}{m_{\chi}+m_{I}}\frac{\rho_I}{\rho_{b}} \int d^{3}\mathbf{v}_{\chi}f_{\chi}  \int d^{3}\mathbf{v}_{I}f_{I}\times\left[\sigma^{I}_{T}\mid\mathbf{v}_{\chi}-\mathbf{v}_{I}\mid \frac{\mathbf{V}_{\chi b}}{V_{\chi b}}\cdot(\mathbf{v}_{\chi}-\mathbf{v}_{I})\right], \nonumber\\
\dot{Q}_{b}&=&\sum_{I}\frac{\rho_{\chi} x_{I}m_{I}}{m_{\chi}+m_{I}} \int d^{3}\mathbf{v}_{\chi}f_{\chi}  \int d^{3}\mathbf{v}_{I}f_{I}\times\left[\sigma^{I}_{T}\mid\mathbf{v}_{\chi}-\mathbf{v}_{I}\mid\mathbf{v}_{\rm{CM}}\cdot (\mathbf{v}_{I}-\mathbf{v}_{\chi})\right], \nonumber\\
\dot{Q}_{\chi}&=&\sum_{I}\frac{\rho_{I}m_{\chi}}{m_{\chi}+m_{I}} \int d^{3}\mathbf{v}_{\chi}f_{\chi}  \int d^{3}\mathbf{v}_{I}f_{I}\times\left[\sigma^{I}_{T}\mid\mathbf{v}_{\chi}-\mathbf{v}_{I}\mid\mathbf{v}_{\rm{CM}}\cdot (\mathbf{v}_{\chi}-\mathbf{v}_{I})\right], 
\end{eqnarray}
where ``dot" refers to derivative over time, the sum is over $e$ and $p$, $\sigma^{I}_{T}$ is given by Eq.(\ref{comp}), 
$\rho_b$, $\rho_I$ and $\rho_{\chi}$ is the baryon, $I$-component of baryon gas and DM density respectively,
$f_{\chi}$ and $f_{I}$ is the phase space density of DM and $I$-component particle respectively, 
$x_{I}$ is the fraction of $I$-component number density, 
and $\mathbf{v}_{i}$ and $\mathbf{V}_{j}$ represent the velocity of particle $i$ and fluid $j$ respectively.
Under this notation, $\mathbf{v}_{\rm{CM}}=(m_{I}\mathbf{v}_{I}+m_{\chi}\mathbf{v}_{\chi})/(m_{I}+m_{\chi})$ is the center-of-mass velocity,
while $V_{\chi b}=\mid \mathbf{V}_{\chi b}\mid=\mid \mathbf{V}_{\chi}-\mathbf{V}_{b}\mid$ is the relative velocity of the two relevant fluids.

Inserting Eq.(\ref{comp}) into Eq.(\ref{DQ1})  and integrating over the particle velocities $\mathbf{v}_{\chi}$ and $\mathbf{v}_{I}$ give rise to the explicit expressions,
\begin{eqnarray}{\label{DQs1}}
D&=&\sum_{I}\hat{\sigma}^{I}\frac{(\rho_{\chi}+\rho_{b})}{m_{\chi}+m_{I}}\frac{\rho_I}{\rho_{b}}\frac{F(r_{I})}{V^{2}_{\chi b}}, \nonumber\\
\dot{Q}_{b}&=&\sum_{I}\frac{\rho_{\chi} x_{I}m_{I}}{(m_{\chi}+m_{I})^{2}}\frac{\hat{\sigma}^{I}}{u_{th,I}}\left[\sqrt{\frac{2}{\pi}}\frac{e^{-r^{2}_{I}/2}}{u^{2}_{th,I}}(T_{\chi}-T_{b})+m_{\chi}\frac{F(r_{I})}{r_{I}}\right], \nonumber\\
\dot{Q}_{\chi}&=&\sum_{I}\frac{\rho_{I}m_{\chi}}{(m_{\chi}+m_{I})^{2}}\frac{\hat{\sigma}^{I}}{u_{th,I}}\left[\sqrt{\frac{2}{\pi}}\frac{e^{-r^{2}_{I}/2}}{u^{2}_{th,I}}(T_{b}-T_{\chi})+m_{I}\frac{F(r_{I})}{r_{I}}\right],
\end{eqnarray}
where $F(r_{I})=\rm{erf}(r_{I}/\sqrt{2})-\sqrt{\frac{2}{\pi}}r_{I}e^{-r_{I}^{2}/2}$ with
\begin{eqnarray}{\label{u}}
r_{I}&=& \frac{V_{\chi b}}{u_{th,I}}, \nonumber\\
u_{th,I}&=&\sqrt{\frac{T_{b}}{m_{I}}+\frac{T_{\chi}}{m_{\chi}}}.
\end{eqnarray}

Taking into account the collision terms, 
one obtains the complete Boltzmann equations \cite{Munoz:2015bca}
for the temperatures of DM and baryon fluid,
\begin{eqnarray}{\label{Beq1}}
\frac{dT_{\chi}}{da}&=& -2 \frac{T_{\chi}}{a} +\frac{2\dot{Q}_{\chi}}{3 aH}, \nonumber\\
\frac{dT_{b}}{da}&=& -2 \frac{T_{b}}{a} + \frac{\Gamma_{C}}{aH}(T_{\gamma}-T_{b})+ \frac{2\dot{Q}_{b}}{3aH}, \nonumber\\
\frac{dV_{\chi b}}{da}&=& - \frac{V_{\chi b}}{a} - \frac{D}{aH}, \nonumber\\
\frac{dx_{e}}{da}&=& -\frac{\mathcal{C}}{aH}\left[n_{H}\mathcal{A}_{B}x^{2}_{e}-4(1-x_{e})\mathcal{B}_{B}e^{3E_{0}/4T_{\gamma}}\right],
\end{eqnarray}
where $H$ is the Hubble parameter, $a=(1+z)^{-1}$ is the scale factor,
$T_{\gamma}(z)=T_{0}(1+z)$ is the CMB temperature with its present value $T_{0}=2.725$ K, 
$\Gamma_{C}$ is the Compton interaction rate,
$\mathcal{C}$ is the Peebles factor \cite{Peebles:1968ja},
$E_0$ is the Hydrogen ground energy, 
$\mathcal{A}_{B}$ and $\mathcal{B}_{B}$ \cite{Ali-Haimoud:2010tlj,Ali-Haimoud:2010hou} are effective recombination coefficient and photoionization rate, respectively.
As seen in Eq.(\ref{Beq1}), a decrease in $T_b$ towards to low red shift requires a negative $\dot{Q}_b$,
which is more easily obtained in small $m_{\chi}$ range as shown in Eq.(\ref{DQs1}).
It is easy to verify that the derived analytical results in Eq.(\ref{DQ1})-(\ref{Beq1}) reduce to those of Refs.\cite{Munoz:2015bca,Tashiro:2014tsa,Dvorkin:2013cea} by taking $I$ as $e$ or $p$.

We will numerically solve Eq.(\ref{Beq1}) via the following initial conditions 
\begin{eqnarray}{\label{ini}}
T_{\chi}(z_{\rm{kin}})&=&0, \nonumber\\
T_{b}(z_{\rm{kin}})&=&T_{\gamma}(z_{\rm{kin}}),  \nonumber\\
x_{e}(z_{\rm{kin}})&\approx& 0.08,
\end{eqnarray}
at the kinetic decoupling with the redshift  $z_{\rm{kin}}=1010$.

\subsection{Two-component DM}
\label{B2}
Now we consider DM composed of two different components $\chi_{i}$, with $i=$1-2.
Without loss of generality, 
we couple the $\chi_{1}$- and $\chi_{2}$-component to the electron and proton of baryon fluid respectively.
Compared to one-component case, what is the same is that there is only a baryon fluid velocity $\mathbf{V}_b$;  what differs is that there are two $\chi_i$ fluid velocities $\mathbf{V}_{\chi_{i}}$. 
As a result, in the two-component case we have two relative fluid velocities $V_{\chi_{i}b}=\mid \mathbf{V}_{\chi_{i}}-\mathbf{V}_{b}\mid$, 
two drag terms $D_{i}=-\dot{V}_{\chi_{i} b}$, and four heating terms $Q_{b_{i}}$ and $Q_{\chi_{i}}$ .

In the same spirit of Eq.(\ref{DQ1}) we have the explicit forms of $D_1$, $Q_{\chi_{1}}$ and  $Q_{b_{1}}$ as follows
\begin{eqnarray}{\label{DQ2}}
D_{1}&=&\hat{\sigma}^{e}\frac{\rho_{\chi_{1}}+\rho_{e}}{m_{\chi_{1}}+m_{e}}\frac{F(r_{e})}{V^{2}_{\chi_{1}b}}+
\hat{\sigma}^{p}\frac{\rho_{\chi_{2}}}{m_{\chi_{2}}+m_{p}}\frac{F(r_{p})}{V^{2}_{\chi_{2}b}}, \nonumber\\
\dot{Q}_{b_{1}}&=&\frac{\rho_{\chi_{1}} x_{e}m_{e}}{(m_{\chi_{1}}+m_{e})^{2}} \frac{\hat{\sigma}^{e}}{u_{th, e}}
\left[\sqrt{\frac{2}{\pi}}\frac{e^{-r^{2}_{e}/2}}{u^{2}_{th,e}}(T_{\chi_{1}}-T_{b})+m_{\chi_{1}}\frac{F(r_{e})}{r_{e}}\right],  \nonumber\\
\dot{Q}_{\chi_{1}}&=&\frac{\rho_{e}m_{\chi_{1}}}{(m_{\chi_{1}}+m_{e})^{2}} \frac{\hat{\sigma}^{e}}{u_{th,e}}
\left[\sqrt{\frac{2}{\pi}}\frac{e^{-r^{2}_{e}/2}}{u^{2}_{th,e}}(T_{b}-T_{\chi_{1}})+m_{e}\frac{F(r_{e})}{r_{e}}\right],
\end{eqnarray}
with 
\begin{eqnarray}{\label{uI}}
r_{e}&=& \frac{V_{\chi_{1}b}}{u_{th,e}}, \nonumber\\
u_{th,e}&=&\sqrt{\frac{T_{b}}{m_{e}}+\frac{T_{\chi_{1}}}{m_{\chi_{1}}}}.
\end{eqnarray}
The forms of $D_2$, $Q_{b_{2}}$ and $Q_{\chi_{2}}$ are obtained by simultaneously replacing $1\rightarrow 2$ and $e\rightarrow p$ in Eqs.(\ref{DQ2})-(\ref{uI}).
Note that for $D_1$ in Eq.(\ref{DQ2}) the second term arises from the $\chi_{2}$-baryon interaction regardless of whether the $\chi_{1}$-baryon interaction is present, and vice versa for $D_2$.
This is one of the key features different from Eq.(\ref{DQs1}) in the previous case.

Equipped with Eq.(\ref{DQ2}),  the Boltzmann equations in Eq.(\ref{Beq1}) are replaced by
\begin{eqnarray}{\label{Beq2}}
\frac{dT_{\chi_{i}}}{da}&=& -2 \frac{T_{\chi_{i}}}{a} +\frac{2\dot{Q}_{\chi_{i}}}{3 aH}, \nonumber\\
\frac{dT_{b}}{da}&=& -2 \frac{T_{b}}{a} + \frac{\Gamma_{C}}{aH}(T_{\gamma}-T_{b})+ \frac{2\sum_{i}\dot{Q}_{bi}}{3aH}, \nonumber\\
\frac{dV_{\chi_{i} b}}{da}&=& - \frac{V_{\chi_{i} b}}{a} - \frac{D_{i}}{aH}, \nonumber\\
\frac{dx_{e}}{da}&=& -\frac{\mathcal{C}}{aH}\left[n_{H}\mathcal{A}_{B}x^{2}_{e}-4(1-x_{e})\mathcal{B}_{B}e^{3E_{0}/4T_{\gamma}}\right],
\end{eqnarray}
where $T_{\chi_{i}}$ refer to the separate temperatures of $\chi_i$ fluids.

Instead of Eq.(\ref{ini}), the initial conditions to Eq.(\ref{Beq2}) are given by
\begin{eqnarray}{\label{ini2}}
T_{\chi_{i}}(z_{\rm{kin}})&=&0, \nonumber\\
T_{b}(z_{\rm{kin}})&=&T_{\gamma}(z_{\rm{kin}}),  \nonumber\\
x_{e}(z_{\rm{kin}})&\approx& 0.08.
\end{eqnarray}

As one derives $T_{21}$ from the Boltzmann equations either in Eq.(\ref{Beq1}) or Eq.(\ref{Beq2}) with an initial value of $V_{\chi b}(z_{\rm{kin}})=V_{\chi b,0}$, 
it is actually described by the Maxwell-Boltzmann distribution 
 \begin{eqnarray}{\label{distribution}}
\mathcal{P}(V_{\chi b,0})=\frac{4\pi }{(2\pi V^{2}_{\rm{rms}}/3)^{3/2}}V^{2}_{\chi b,0}e^{-3V^{2}_{\chi b,0}/(2V^{2}_{\rm{rms}})},
\end{eqnarray}
with $V_{\rm{rms}}\approx 29$ km/s at the kinetic decoupling.
So, the final value of  $T_{21}$ should be sky-averaged as follows
\begin{equation}{\label{avg}}
\left<T_{21}\right>=
 \left\{
\begin{array}{lcl}
\int dV_{\chi b} \mathcal{P}(V_{\chi b,0})T_{21}(V_{\chi b,0}), ~~~~~~~~~~~~~~~~~~~~~~~~~~~~~~~~ \rm{one-component},\\
\int dV_{\chi_{1} b}dV_{\chi_{2} b}\mathcal{P}(V_{\chi_{1} b,0})\mathcal{P}(V_{\chi_{2} b,0})T_{21}(V_{\chi_{1} b,0}, V_{\chi_{2} b,0}), ~~ \rm{two-component},\\
\end{array}\right.
\end{equation}
where each of $V_{\chi_{i} b,0}$ in the two-component case satisfies the same Maxwell-Boltzmann distribution in Eq.(\ref{distribution}).
The thermal average in Eq.(\ref{avg}) is necessary in order to eliminate statistical errors, 
even though this demands a larger compute source to handle numerical analysis in the next section.

\section{Results}
\label{space}
Let us now discuss the parameter spaces which can explain the observed EDGES 21-cm signal. 
We divide our study into two representative cases as shown in Table.\ref{cases},
where the one-component DM contains only a force carrier ($\omega$) and the two-component DM contains two different force carriers $\omega_{i}$ respectively. 
In the first case it is sufficient to introduce the DM mass $m_{\chi}$ and two cross sections $\hat{\sigma}^{e}$ and $\hat{\sigma}^{p}$ to parametrize the parameter space,
while in the later case we introduce two DM masses $m_{\chi_{i}}$, two cross sections as above, and a new parameter $\delta$
\begin{eqnarray}{\label{delta}}
\rho_{1}&=&\delta\rho_{\rm{CDM}}, \nonumber\\
\rho_{2}&=&(1-\delta)\rho_{\rm{CDM}},
\end{eqnarray}
to describe the fraction of each DM component energy density,
where $ \rho_{\rm{CDM}}\approx 0.3$ GeV$/$cm$^{3}$ is the observed cold DM energy density.

\begin{table}
  \centering
\begin{tabular}{|c|c|c|}
 \hline
  &  Pattern of DM-baryon interaction  & Parameters \\  \hline
 one-component &~~~~~~\begin{minipage}[b]{0.3\columnwidth}
		\centering
		\raisebox{-0.7\height}{\includegraphics[width=\linewidth]{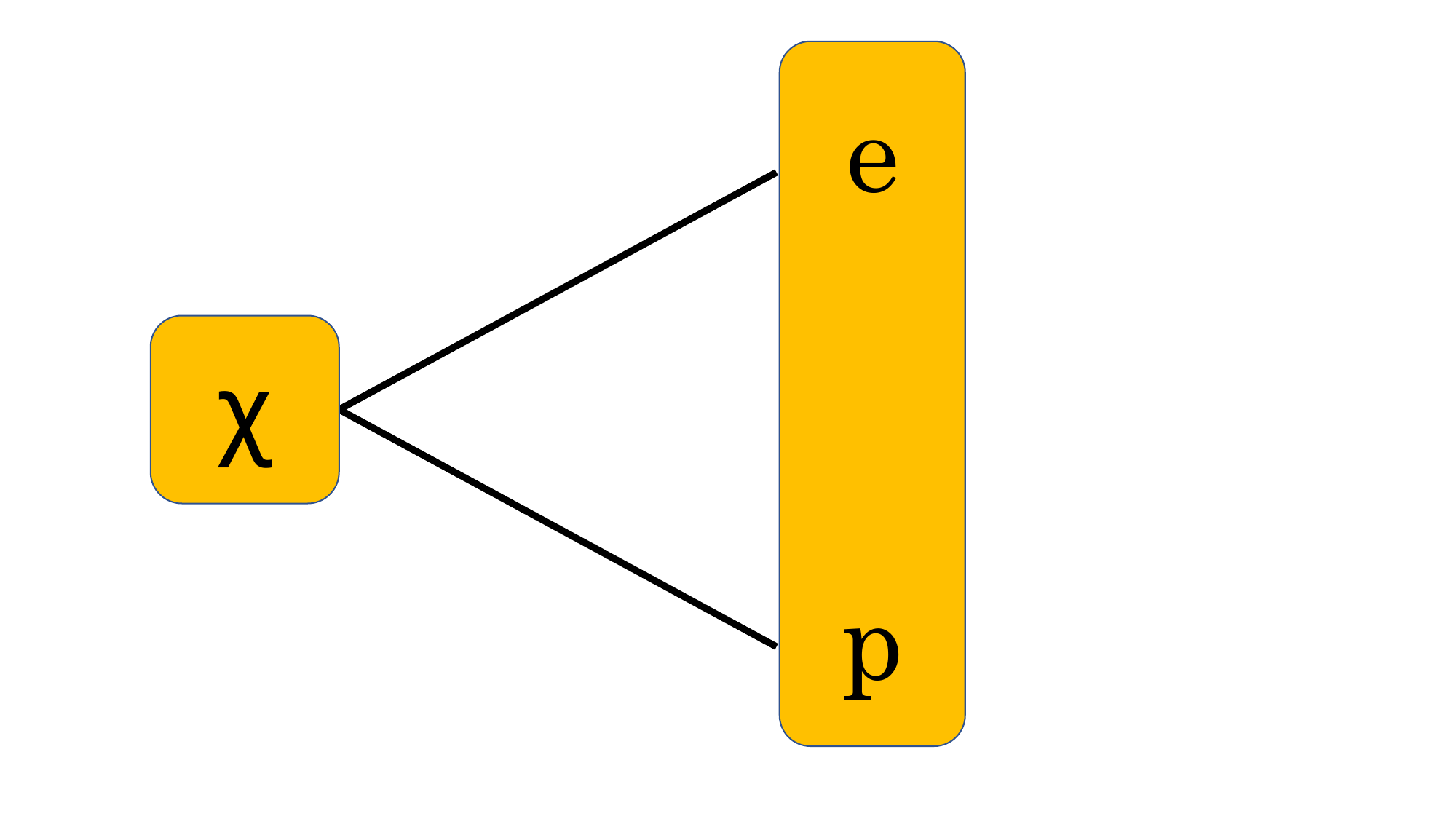}}
	\end{minipage} 
    & $m_{\chi}, \hat{\sigma}_{e}, \hat{\sigma}_{p}$ \\  \hline
two-component  &\begin{minipage}[b]{0.3\columnwidth}
		\centering
		\raisebox{-0.7\height}{\includegraphics[width=\linewidth]{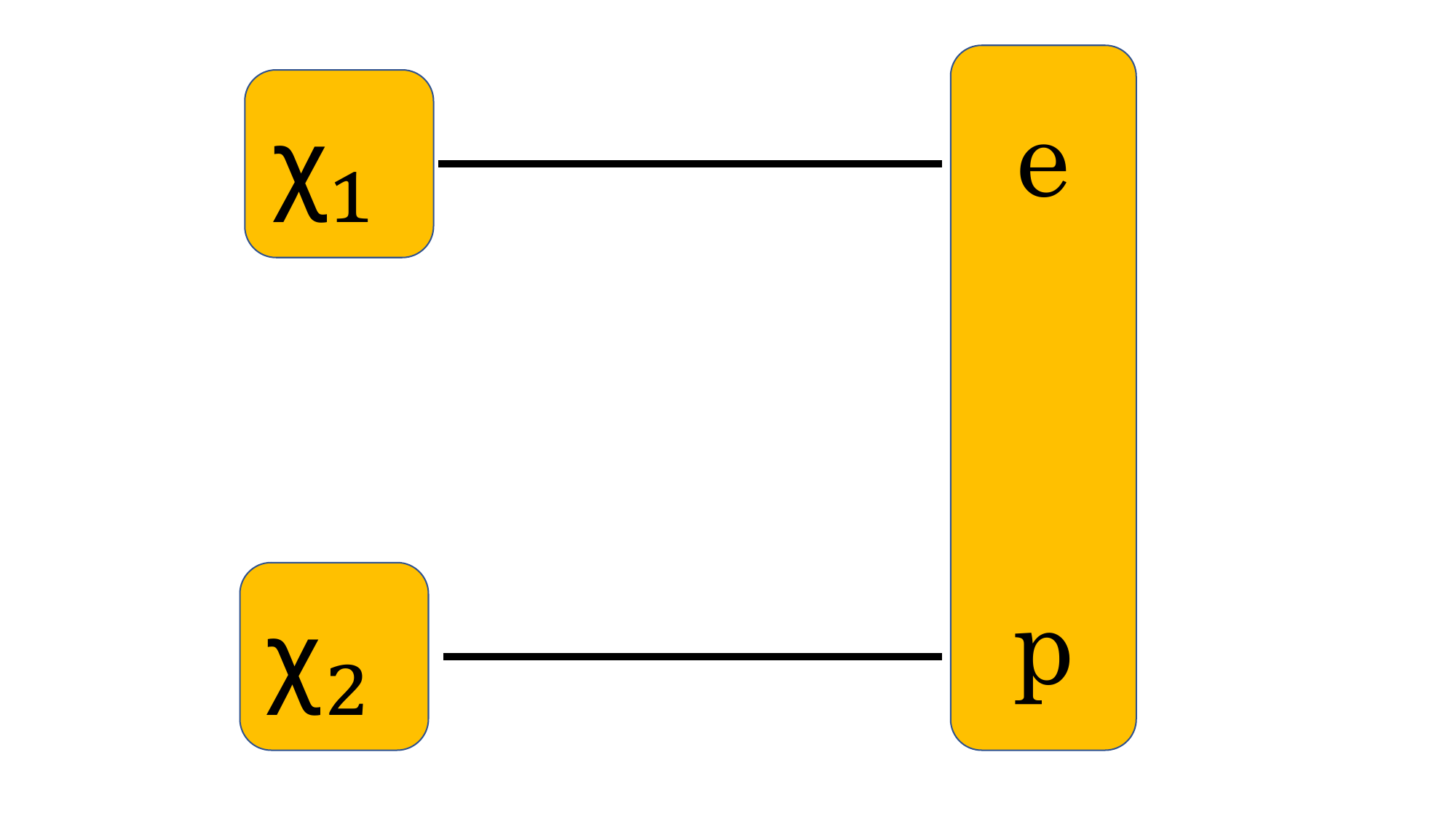}}
	\end{minipage}
    & $m_{\chi_{i}}, \hat{\sigma}_{e}, \hat{\sigma}_{p},\delta$ \\ \hline
\end{tabular}
\caption{Two DM scenarios considered. In the one-component case, DM $\chi$ is assumed to couple to both $e$ and $p$ via a light force carries $\omega$ with mass in the range of eV-keV. Together with $m_{\chi}$, the two scattering cross sections $\sigma^{I}_{T}$ with $I=\{e,p\}$ which scale as in Eq.(\ref{comp}) are used as the input parameters. Likewise, in the two-component case DM is composed of two different fields $\chi_{1}$ and $\chi_{2}$ coupled to $e$ via $\omega_{1}$ and $p$ via $\omega_{2}$ respectively, with the masses of two force carries in the range of eV-keV, two cross sections $\sigma^{I}_{T}$ scaling as in Eq.(\ref{comp}), and a fraction of DM energy density denoted by $\delta$.}
\label{cases}
\end{table}

\subsection{One-component DM}
\label{space1}
\begin{figure}
\centering
\includegraphics[width=15cm,height=9cm]{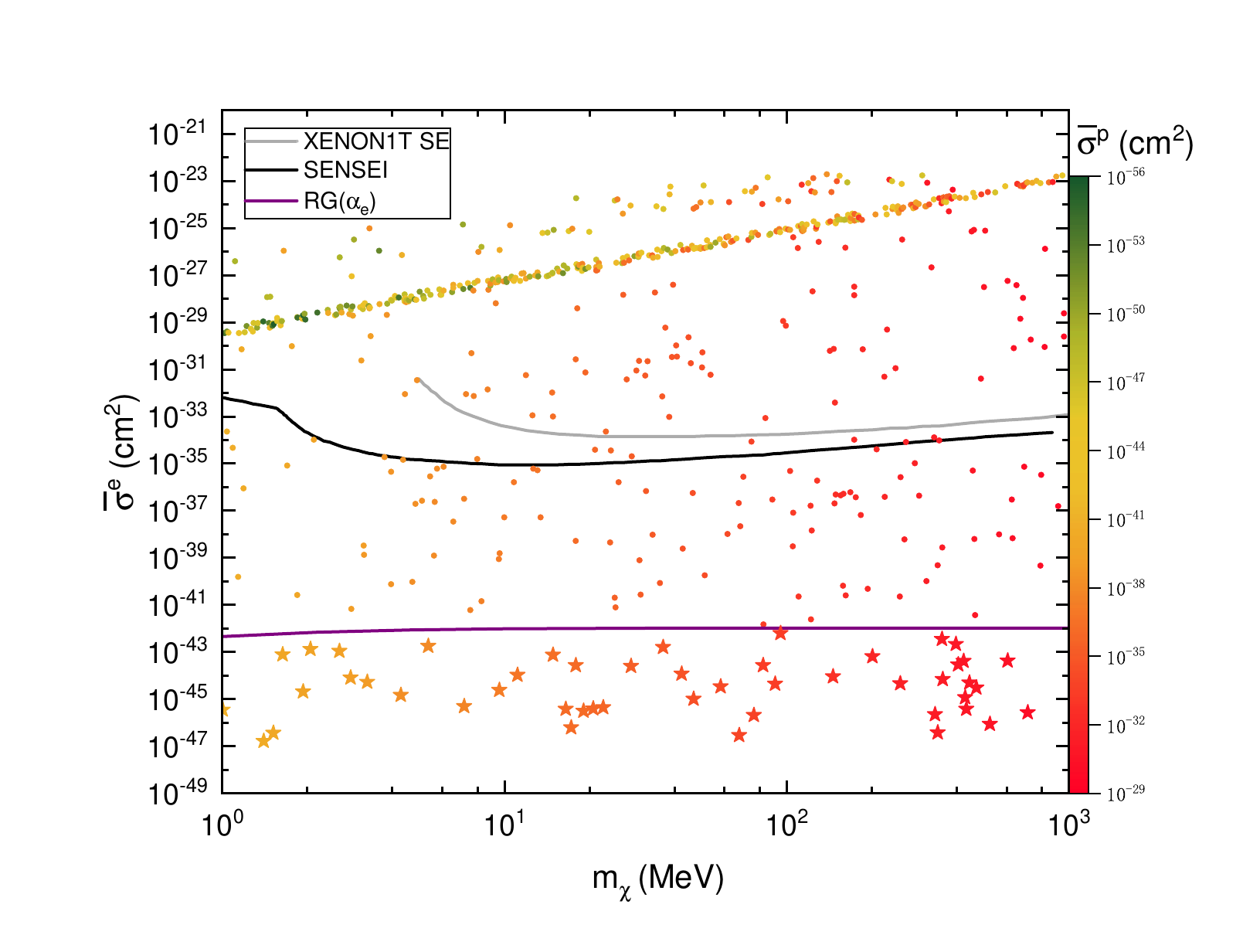}
\includegraphics[width=15cm,height=9cm]{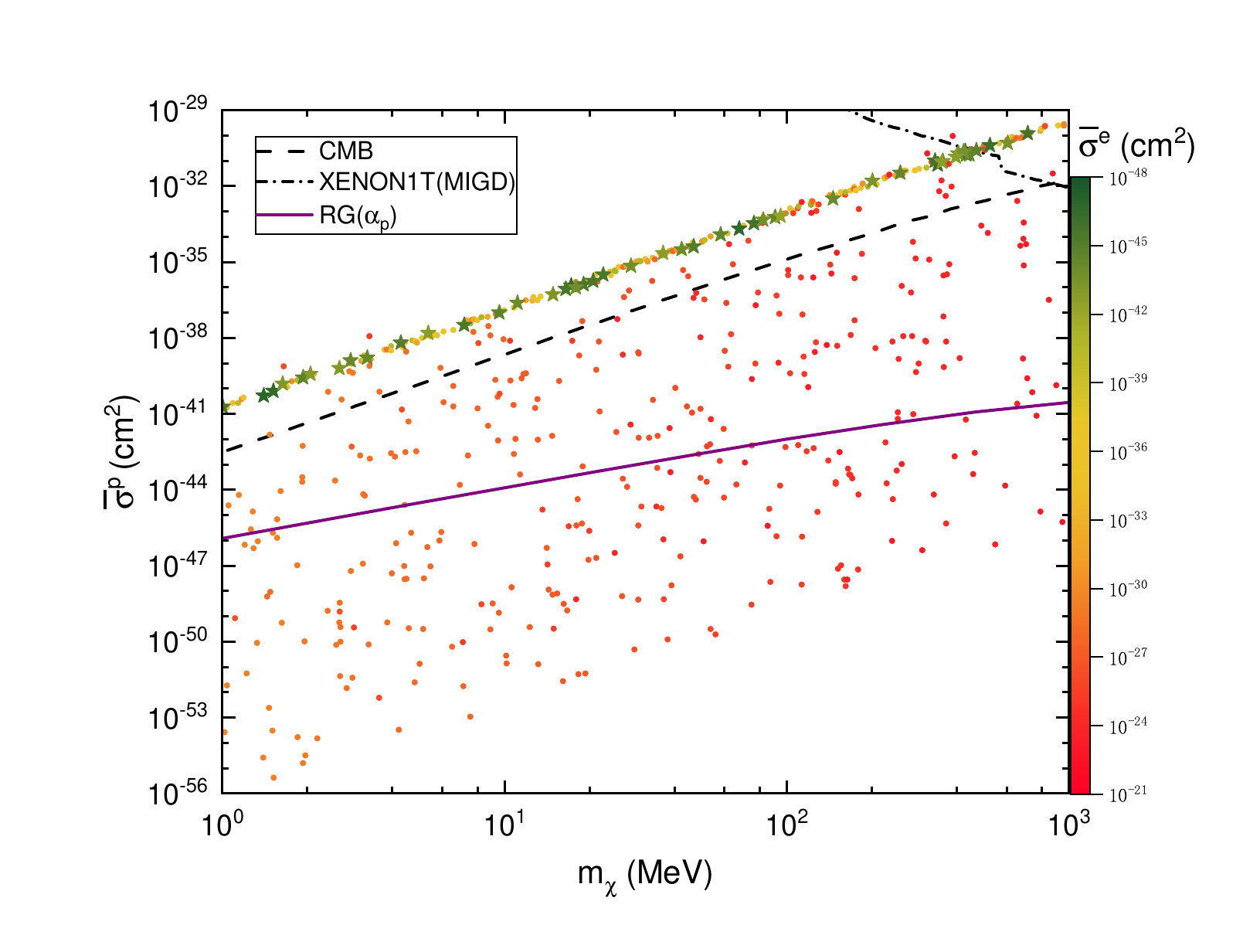}
\centering
\caption{Samples which yield $-500 \leq \left<T_{21}\right> \leq -300$ mK reported by the EDGES experiment in the one-component DM with $m_{\omega}=1$ eV, 
which are projected to the plot of $m_{\chi}-\bar{\sigma}^{e}$ ($\mathbf{top}$) and $m_{\chi}-\bar{\sigma}^{p}$ ($\mathbf{bottom}$) respectively.
For comparison, we show the SENSEI limit (black) \cite{SENSEI:2020dpa} on DM-e scattering, 
the XENON1T limits on DM-e (gray) \cite{XENON:2021myl} and  spin-dependent DM-p (dotted-dashed) \cite{XENON:2019zpr} scattering,
the CMB limit (dashed) \cite{Xu:2018efh} on DM-p scattering, 
and the stellar cooling limits \cite{Hardy:2016kme} (purple) simultaneously. 
Samples above the limits are excluded. 
See text for details.}
\label{odm}
\end{figure}

Fig.\ref{odm} shows the parameter space of the one-component DM model which satisfies $-500 \leq T_{21}\leq -300$ mK as reported by the EDGES experiment with $m_{\omega}=1$ eV.
Note the mass range of $m_{\omega}$ allowed by the requirement $m_{\omega}\leq 10^{-6}\mu_{I}$  in the whole DM range $m_{\chi}\sim 1-10^{3}$ MeV for both $I=e$ and $I=p$ is at most of order $\sim$ eV scale.
For each sample in this figure, we have chosen $q_{p}\approx 2$ MeV in converting $\hat{\sigma}^{p}$ to $\bar{\sigma}^{p}$ in Eq.(\ref{transf}),
which seeds an uncertainty of order $\sim$ a few times in $\bar{\sigma}^{p}$ in certain DM mass range.

To identify whether a constraint is model-independent or model-dependent, 
we remind that behind the cross sections 
\begin{eqnarray}{\label{ecoupling}}
\bar{\sigma}^{I}\approx \frac{16\pi\alpha_{I}\alpha_{\chi}}{q^{4}_{I}}\mu^{2}_{I},
\end{eqnarray}
there are two structure constants $\alpha_{I}$ and $\alpha_{\chi}$ with respect to $\omega$-SM and $\omega$-DM system respectively, see Fig.\ref{sketch} for a sketch.
Let us examine the constraints mentioned in Sec.\ref{introduction} one by one.
\begin{itemize}
\item DM direct detections. 
These limits are model-independent.
In the mass range of $m_{\chi}\sim 1-10^{3}$ MeV,
the most stringent limits on $\bar{\sigma}^{e}$ arise from the SENSEI (black) \cite{SENSEI:2020dpa} and XENON1T (gray)\cite{XENON:2021myl}),
while up-to-date spin-dependent limit on $\bar{\sigma}^{p}$ from the XENON1T (dotted-dashed) \cite{XENON:2019zpr}.
\item BBN constraint. BBN places a constraint on effective number of neutrinos $N_{\rm{eff}}$ 
due to relativistic effect \cite{Vogel:2013raa} of the light $\omega$.
Since the effect on $N_{\rm{eff}}$ is determined by the mediator energy density $\rho_{\omega}$, 
the BBN constraint is model-dependent.
\item CMB constraint. Similar to the BBN constraint, the CMB constraint on $N_{\rm{eff}}$ \cite{Vogel:2013raa} is also model-dependent. 
Apart from $N_{\rm{eff}}$, measurements on CMB anisotropy offer another method to precisely constrain the DM-baryon interactions. A model-independent upper bound on $\bar{\sigma}^{p}$ (dashed) can be found in \cite{Xu:2018efh} without the DM-e interaction (i.e. $\alpha_{e}=0$). 
\item Supernova 1987A (SN1987A). 
The energy loss of SN1987A into both $\chi$ and $\omega$ particles places an upper bound dependent on $\alpha_{\chi}$, $\alpha_{e}$ and $\alpha_{p}$.
So far, the available SN1987A bounds in the literature \cite{Chang:2018rso,Hardy:2016kme} only apply to specific DM models such as dark photon DM model with $\alpha_{e}\sim \alpha_{p}$.
\item Stellar cooling. In a stellar such as the sun, horizontal-branch stars or red-giants (RG), 
the energy loss to the light $\omega$ particles with the mass scale $m_{\omega}$ smaller than keV \cite{Hardy:2016kme, An:2014twa} can be significant. 
By turning off $\alpha_{p}$ ($\alpha_{e}$),  
the stellar cooling bound\footnote{A stellar system provides a local thermal bath where a large number of the $\omega$ particles can be produced. This contributes to a new stellar energy loss after the produced particles escape the core of the stellar system as a result of the feeble interactions with the thermal bath therein. 
The stellar cooling bounds are derived as follows. 
(i) Calculate the squared amplitude of relevant processes.  Here, the main processes include Bremsstrahlung and Compton emission of $\omega$. (ii) Integrate over momentum space. It is more convenient to transform the integration over momentums into an integration over dimensionless variables. (iii) Compare the stellar cooling rate to observed limit on the luminosity of a stellar by taking into account relevant stellar parameters such as stellar core temperature and radius.  The derivation becomes more complicated if additional effects such as photon polarization and  dependence of stellar parameters on radius are needed to be considered.} 
on $\alpha_{e}$ ($\alpha_{p}$) \cite{Hardy:2016kme} can be promoted to model-independent constraints with the help of a rational bound $\alpha_{\chi}\leq 1$, 
as shown in the plot of $m_{\chi}-\bar{\sigma}^{e}(\bar{\sigma}^{p})$ in purple.
\item LSS.  This imposes a rough upper bound  \cite{Tulin:2013teo} on effect of DM self interaction controlled by $\alpha_{\chi}$, which can be satisfied by $\alpha_{\chi}\leq 1$.
\item Colliders. The constraints on $\alpha_e$ by lepton colliders such as LEP and on $\alpha_p$ by hadron colliders such as LHC are obviously less competitive than the stellar cooling bounds in the parameter regions considered.
\end{itemize}
Compared to the DM direct detection, BBN, CMB and collider limits,
the stellar cooling bounds are used for estimates of magnitudes of the structure constants at best.
Even so, they are the most stringent constraints in the parameter regions where they are present.
This point can be easily verified by choosing an explicit value of $\alpha_{e}$ or $\alpha_{p}$ below the stellar cooling bounds.

Fig.\ref{odm} reveals two key points.
The first point is that the parameter space for the simplified one-component DM models where either $\alpha_{e}$ or $\alpha_{p}$ is absent 
corresponds to the parameter regions with negligible  $\bar{\sigma}^{e}$ or $\bar{\sigma}^{p}$ respectively. 
To manifest this, we label the samples in the case with tiny $\bar{\sigma}^{e}$ by ``$\star$".
While below the stellar cooling bound on $\alpha_e$ in the $\mathbf{top}$ plot,
these star points are excluded by the stellar cooling bound on $\alpha_p$ in the $\mathbf{bottom}$ plot,
and vice versa. This result applies to simplified one-component freeze-in  DM model with the light force carrier as a gauged $L_{e}-L_{\mu}$ or $L_{e}-L_{\tau}$ \cite{He:1990pn, Ma:2001md, Bauer:2018onh,Wise:2018rnb}.
The second point is that when $\alpha_{e}$ and $\alpha_{p}$ are present and their contributions to $\left<T_{21}\right>$ are comparable,
the samples are clearly excluded by the stellar cooling bounds if not by the DM direct detection limits etc.
This result applies to simplified one-component freeze-in  DM with the light force carrier as a gauged $B-L$ \cite{Heeck:2014zfa,Bilmis:2015lja,Ilten:2018crw,Bauer:2018onh} with $\alpha_{e}/\alpha_{p}\sim 1$, 
dark photon \cite{Holdom:1985ag, Boehm:2003hm,Pospelov:2007mp} with $\alpha_{e}/\alpha_{p}\sim 1$,
or axion-like particle (ALP)\footnote{For recent reviews, see \cite{Marsh:2015xka,DiLuzio:2020wdo,Sikivie:2020zpn}. 
The spin-dependent couplings of ALP to the SM fermions guarantee that the interplay between the DM and hydrogen atom of the baryon gas is negligible.} with various ratios of $\alpha_{e}/\alpha_{p}$.

The exclusion is robust because $i)$ when both $\alpha_{e}$ and $\alpha_{p}$ are present the stellar cooling bounds are still valid as an estimate of order of magnitudes, 
and $ii)$ the magnitudes of the cross sections for the samples in Fig.\ref{odm} change at most by one-to-two orders 
when the value of $m_{\omega}$ is adjusted in the allowed mass range of 1-$10^3$ eV.
These results partially explain the motivation to propose the mini-charged DM model as mentioned in the Introduction, 
where the stringent stellar cooling bounds no longer exist in the DM mass range with $m_{\chi}$ above MeV scale.

\subsection{Two-component DM}
\label{space2}
Similar to the one-component DM, 
we now present the parameter space of two-component  DM which can explain the EDGES data.
In this situation, $\bar{\sigma}^{I}$ depend on two different masses $m_{\omega_{i}}$ instead of one in the one-component DM. 
So there are more options on $m_{\omega_{i}}$ to satisfy the requirements $m_{\omega_{1}}\leq 10^{-6}\mu_{e}$ and $m_{\omega_{2}}\leq 10^{-6}\mu_{p}$.
In the following we consider two specific cases: a) $m_{\omega_{1}}=m_{\omega_{2}}=1$ eV, and b) $m_{\omega_{1}}=1$ eV, $m_{\omega_{2}}=10^{2}$ eV, where the mass difference between $m_{\omega_{1}}$ and $m_{\omega_{2}}$ in the later case is large.
Note that the former case does not reduce to the one-component DM because of the presence of $\delta$. 

\begin{figure}
\centering
\includegraphics[width=15cm,height=9.2cm]{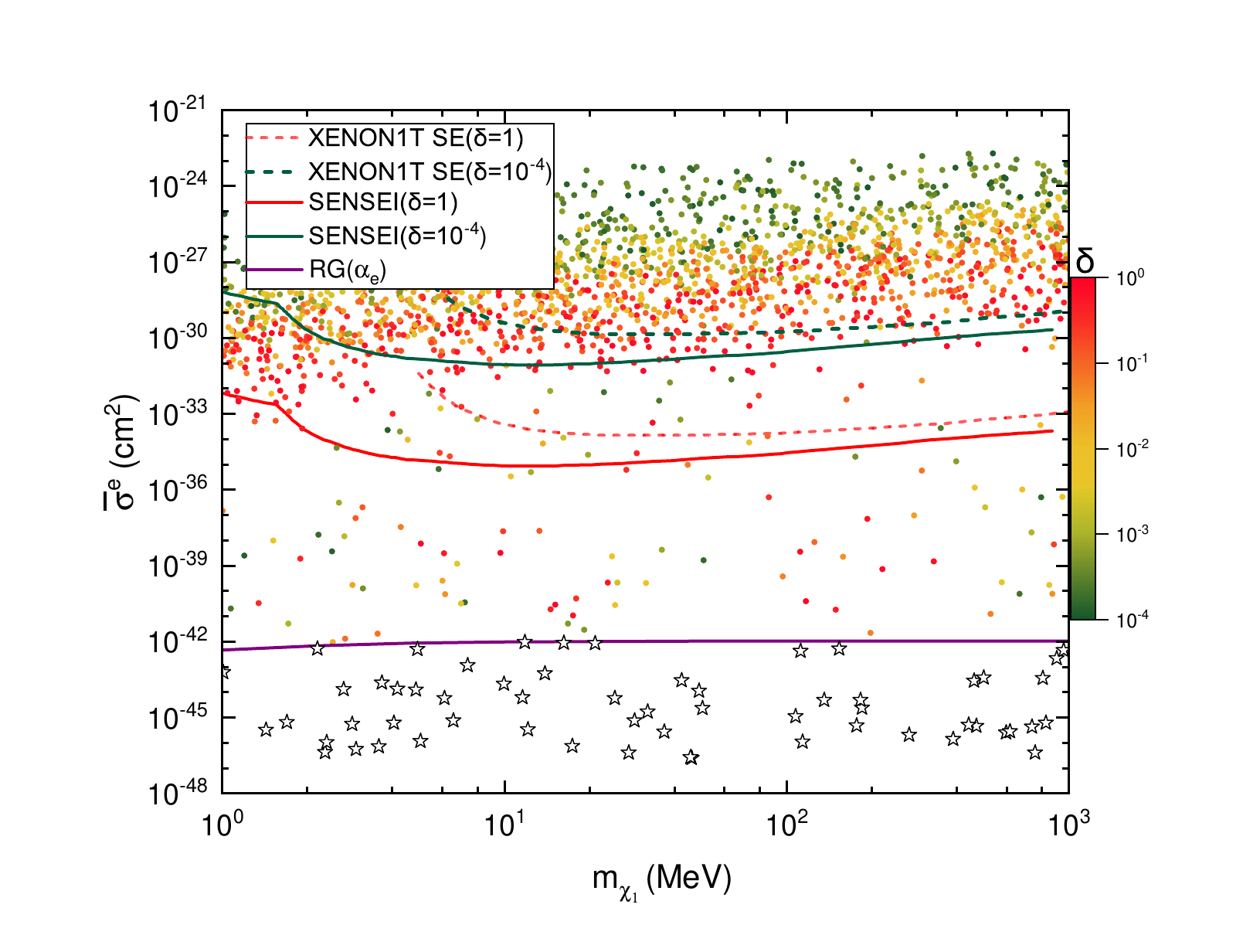}
\includegraphics[width=15cm,height=9.2cm]{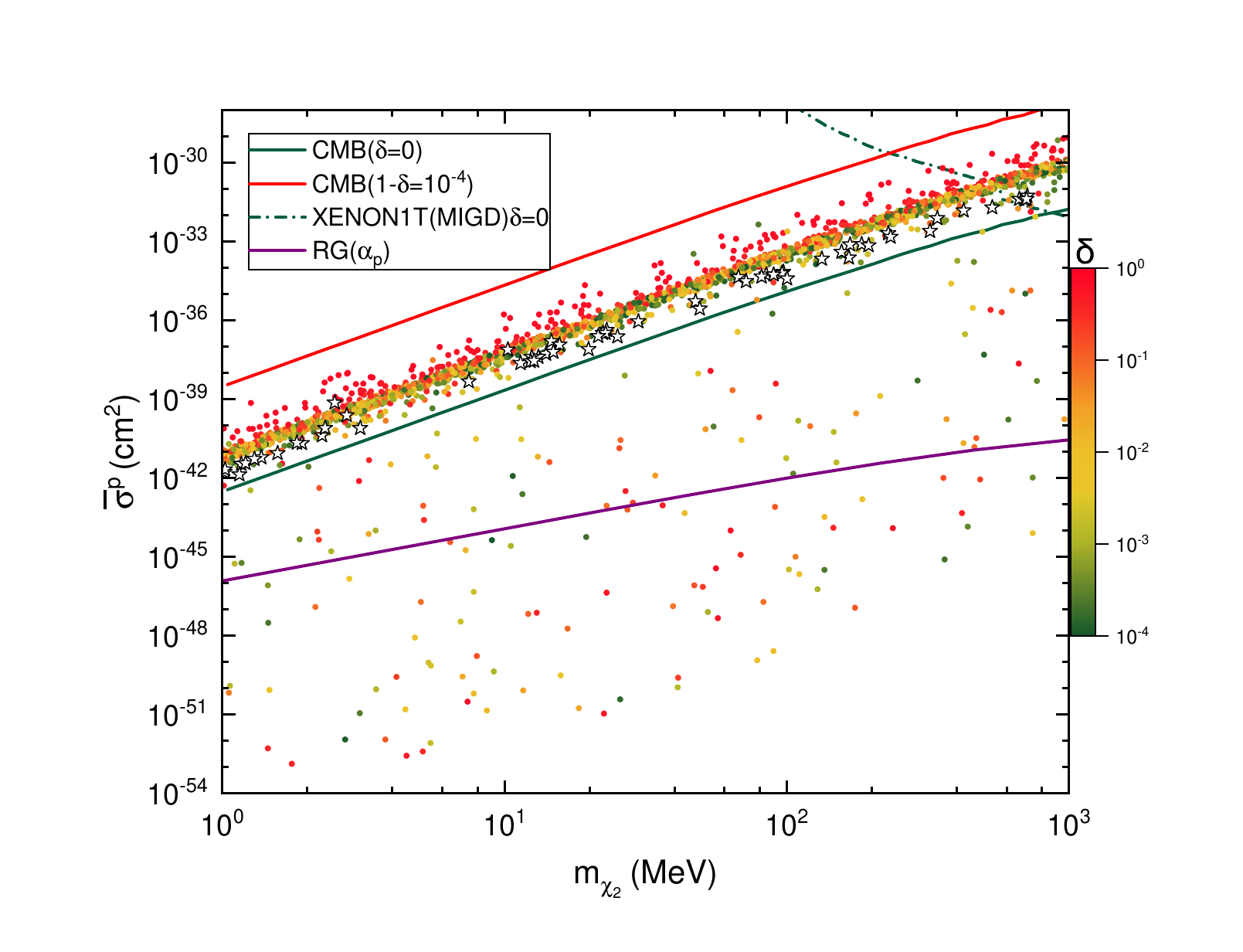}
\centering
\caption{Samples which yield $-500 \leq \left<T_{21}\right> \leq -300$ mK reported by the EDGES experiment in the two-component DM with $m_{\omega_{1}}=m_{\omega_{2}}=1$ eV, 
which are projected to the plot of $m_{\chi_{1}}-\bar{\sigma}^{e}$ ($\mathbf{top}$) and $m_{\chi_{2}}-\bar{\sigma}^{p}$ ($\mathbf{bottom}$) respectively, 
with the dependence on the fraction parameter $\delta$ highlighted. 
We also present the same constraints as in Fig.\ref{odm} for comparison, 
with the dependences of the DM direct detection and CMB limits on $\delta$ illustrated for explicit values of $\delta\approx \{0, 10^{-4},1\}$.}
\label{tdm1}
\end{figure}

The model-independent constraints in the two-component DM are the same as in the case of one-component DM,
despite that some of them have to be properly modified. Fortunately, the task is not impossible.
\begin{itemize}
\item DM direct detections. These limits have to be modified,
as for $\delta\neq 0,1$ each DM component $\chi_i$ only constitutes a portion of the observed DM energy density. 
The DM direct detection limits should be rescaled by an overall factor $\delta^{-1}$ or $(1-\delta)^{-1}$ for $\chi_{1}$ and $\chi_{2}$ respectively, as seen from Eq.(\ref{delta}). 
Because the signal rate at the individual DM direct detection experiment is linearly proportional to the number density of each DM component involved.\footnote{This simple rescaling is no longer valid if each DM component $\chi_i$ simultaneously interacts with both $e$ and $p$.}
\item CMB constraint. Following that CMB is a linear cosmology, the fractional difference of the temperature and polarization CMB power spectra due to the DM-baryon interaction is linearly proportional to the DM energy density through Boltzmann equations \cite{Xu:2018efh}.
Therefore, the CMB limit on $\bar{\sigma}^{p}$ for $\chi_{2}$ is obtained via rescaling the original limit by a factor $(1-\delta)^{-1}$.
\item Stellar cooling. Unlike in the one-component DM where $\chi$ couples to electron and proton simultaneously,  
 the stellar cooling bound on $\hat{\sigma}^{e}$ ($\hat{\sigma}^{p}$) is solely  set on $\chi_{1}$ ($\chi_{2}$). 
\end{itemize}

\begin{figure}
\centering
\includegraphics[width=15cm,height=9.2cm]{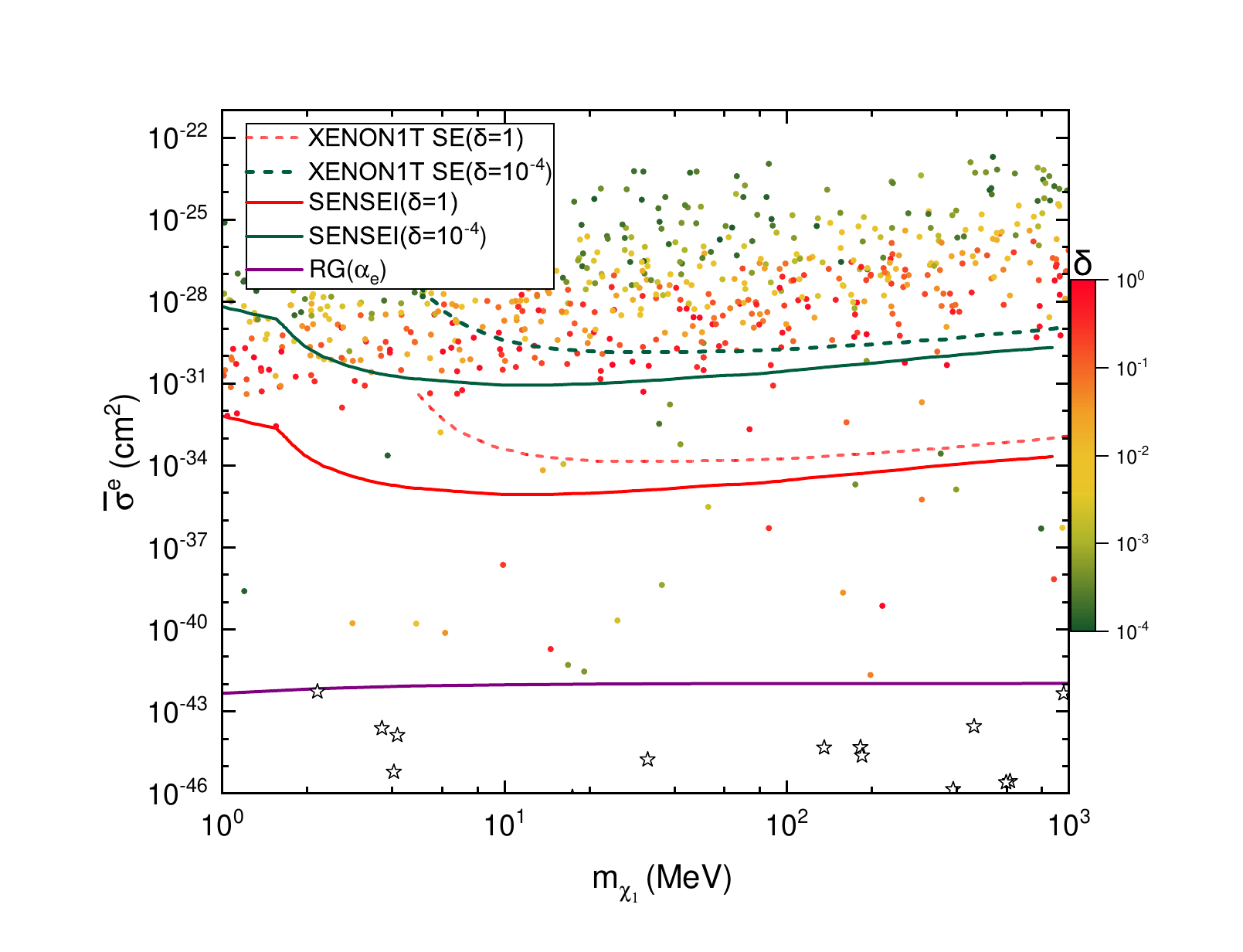}
\includegraphics[width=15cm,height=9.2cm]{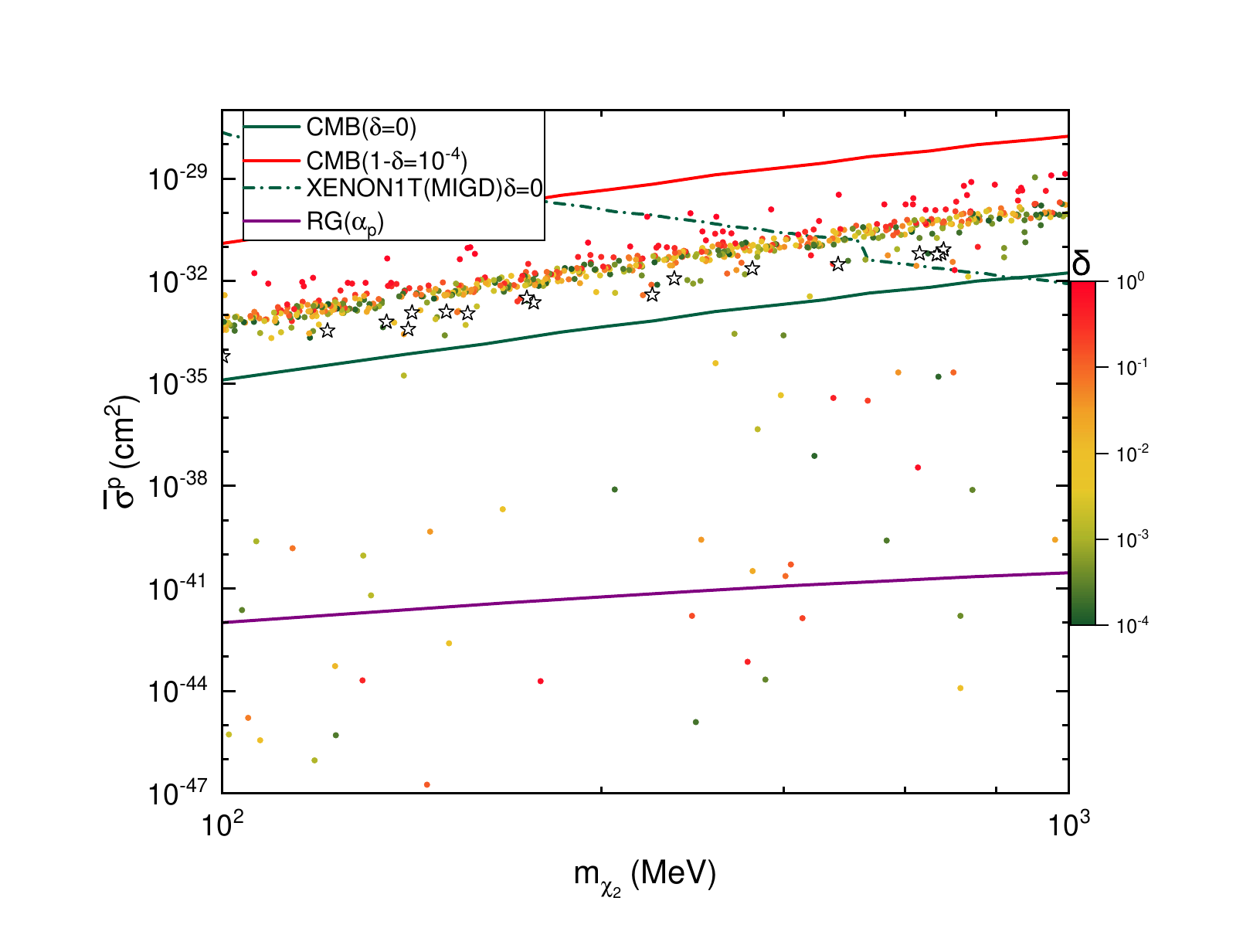}
\centering
\caption{Same as Fig.\ref{tdm1} with $m_{\omega_{1}}=1$ eV and $m_{\omega_{2}}=10^{2}$ eV instead.}
\label{tdm2}
\end{figure}

Fig.\ref{tdm1} shows the samples which gives rise to $-500 \leq \left<T_{21}\right> \leq -300$ mK in the two-component DM with $m_{\omega_{1}}=m_{\omega_{2}}=1$ eV, where the dependence on the fraction parameter $\delta$ is highlighted by the color bar. 
Unlike the DM direct detection limits etc, the stellar cooling bounds do not change, 
as they are independent on $\delta$. 
As the blue points in the previous one-component DM in Fig.\ref{odm}, 
we label the special points below the stellar cooling bound on $\alpha_e$ by ``star" in the $\mathbf{top}$ plot,
all of which turn out to be excluded by the stellar cooling bound on $\alpha_p$ in the $\mathbf{bottom}$ plot. 
Similarly, the samples below the stellar cooling bound on $\alpha_p$ in the $\mathbf{bottom}$ plot are excluded by the stellar cooling bound on $\alpha_e$ in the $\mathbf{top}$ plot, which are not explicitly shown.
Moreover, the exclusion holds even when the force carrier masses are adjusted in their allowed ranges. 
To manifest this point, we show in Fig.\ref{tdm2} the two-component DM with $m_{\omega_{1}}=1$ eV and $m_{\omega_{2}}=10^{2}$ eV.
Compared to Fig.\ref{tdm1}, the number of samples in Fig.\ref{tdm2} is reduced by the stronger requirement $m_{\chi_{2}}\geq 10^{6}m_{\omega_{2}}\sim 10^{2}$ MeV as shown in the $\mathbf{bottom}$ plot.

Similar to the previous one-component case, the exclusion is robust.
This result applies to simplified two-component freeze-in  DM with the two light force carriers 
as two scalars such as two ALPs with spin-dependent couplings to the SM quarks and leptons respectively, among others.

\section{Conclusion}
\label{con}
The brightness temperature of hydrogen 21-cm line reported by the EDGES experiment implies that the temperature of the baryon gas is lower than the prediction of $\Lambda$CDM. 
To cool down the baryon gas after the kinetic decoupling, it is natural to consider DM-baryon interaction.
DM-baryon interactions with magnitudes of the scattering cross sections below current DM direct detection thresholds hardly accommodate the observed signal, unless they are velocity-dependent e.g, in the Coulomb-like interaction.
Previous studies show this type of interaction being still inadequate in simple freeze-in  DM models.
In this study we have performed a model-independent analysis on the parameter space either in one-component and two-component freeze-in  DM with the light force carrier(s) not being photon. 
To achieve this, we have provided necessary background materials such as the conversion of the two different cross sections used by the relevant experiments and the Boltzmann equations which govern the evolution of the temperatures of both DM and baryon fluid.
We have shown that both cases are robustly excluded by the stringent stellar cooling bounds if not by the DM direct detections etc in the sub-GeV DM mass range.
The exclusion of one-component case applies to freeze-in  DM model with the light force carrier as gauged $B-L$, $L_{e}-L_{\mu}$, $L_{e}-L_{\tau}$, dark photon or ALP,
while the exclusion of two-component case applies to freeze-in  DM model with the two light force carriers as two scalars such as two ALPs with spin-dependent couplings to the SM quarks and leptons respectively.
These new results, together with the earlier findings in the literature, 
nearly close the window of (simplified) freeze-in DM with the Coulomb-like interaction with the baryon gas
as a solution to the EDGES 21-cm signal.

\section*{Acknowledgments}
This work is supported in part by the National Natural Science Foundation of China with Grant No. 11775039 and
the High-level Talents Research and Startup Foundation Projects for Doctors of Zhoukou Normal University with Grant No.ZKNUC2021006.

\end{document}